\documentstyle[prb,aps,epsf,preprint]{revtex}
\tightenlines
\begin{document}
\title{Model of charge and magnetic order formation\\ in itinerant electron systems}
\author{Romuald Lema\'{n}ski}
\address{Institute of Low Temperature and Structure Research, \\
Polish Academy of Sciences, Wroc\l aw, Poland}
\maketitle

\begin{abstract}
We propose a simple model of charge and/or magnetic order formation in
systems containing both localized and itinerant electrons coupled by
the on-site, spin-dependent interaction that represents Coulomb
repulsion and Hund's rule (a generalized Falicov-Kimball model).
Ground state properties of the model are analyzed on the square lattice
on a basis of the phase diagrams that have been constructed rigorously,
but in a restricted configurational space. For intermediate values
of the coupling constants there are considerable ranges of itinerant
electron densities where phases with complex charge and magnetic
structures of the localized electrons have lower energy than the
simplest antiferro- and ferromagnetic ones. A strong tendency towards
the antiferromagnetic coupling between spins of localized electrons has
been observed close to half-filling for any density of localized electrons,
including situations where the magnetic ions are diluted. For small band
fillings the ferromagnetic coupling between localized spins is predominant.

\end{abstract}
\section{Introduction}
The purpose of this contribution is to introduce a simple model of charge
and/or magnetic order formation generated by an on-site, spin-dependent
interaction of itinerant electrons with localized ones, and to provide
a preliminary analysis of that model. The model is able to capture many
essential aspects of magnetism of both the localized and itinerant electrons.
It is based on the Falicov-Kimball model (FKM), which was originally
proposed as a model of metal-insulator transitions in mixed-valence
compounds \cite{FalicovKimball}. Later on the FKM appeared to be suitable
for studying some other phenomena as well, as e.g. a tendency for
charge-density-wave formation in interacting fermion systems
\cite{LachLyzwaJedrzejewski,WatsonLemanski,LemanskiFreericksBanach}.
Actually, in most of applications studied so far the spinless version
of the model was explored. As far as we know, spin degrees of freedom
were taken into account only in a few papers
\cite{Brandt,Farkasovsky,FreericksZlatic,ZlaticFreericksLemanskiCzycholl,FreericksZlaticRMP,Jedrzejewski,LemanskiWojtkiewicz},
but in all of them the interactions between electrons were spin-independent.

However, many experiments show that a charge superstructure often occurs
together with a magnetic order \cite{MookDaiDogan,Ando,HowaldEisakiKanekoKapitulnik}.
In order to describe both within a single model we proposed a generalization
of the FKM with an anisotropic, spin-dependent local interaction. These ideas have
also been applied to the Hubbard model \cite{FrohlichUeltschi}, but that work
has been performed only in the limit of a large spin coupling constant;
the authors named this model the "2-band Ising-Hubbard model".
Hence we call our model the Ising-Falicov-Kimball model.

The model Hamiltonian is
\begin{eqnarray}
\label{ham}
H= & \sum\limits_{<m,n>}\sum\limits_{\sigma=\uparrow,\downarrow}
(t-\mu_d\delta_{nm} ) d^+_{m,\sigma} d_{n,\sigma}
+(E_f-\mu_f)\sum\limits_{m}\sum\limits_{\eta=\uparrow,\downarrow}
 f^+_{m,\eta} f_{m,\eta} & \nonumber \\
 & +U \sum\limits_{m}\sum\limits_{\sigma ,\eta =\uparrow,\downarrow}
d^+_{m,\sigma} d_{m,\sigma}
f^+_{m,\eta} f_{m,\eta} & \\
& +J \sum\limits_{m}\sum\limits_{\sigma=\uparrow,\downarrow}
(d^+_{m,\sigma} d_{m,\sigma}
f^+_{m,-\sigma} f_{m,-\sigma}
-d^+_{m,\sigma} d_{m,\sigma}
f^+_{m,\sigma} f_{m,\sigma}), & \nonumber \;
\end{eqnarray}
where $<m,n>$ means the nearest neighbor lattice sites $m$ and $n$,
$\sigma$ and $\eta$ are $spin-1/2$ indices, $d_{m,\sigma}$  ($f_{m,\sigma}$)
and $d^+_{m,\sigma}$ ($f^+_{m,\sigma}$) are annihilation and creation
operators of itinerant (localized) electrons, respectively.
The on-site interaction between localized and itinerant
electrons is represented by two coupling constants: $U$, which is
spin-independent Coulomb-type and $J$, which is spin-dependent Ising-type.
The later reflects the Hund's rule force.
The other parameters of the model are: the hopping amplitude $t$
(we will set it equal to one for our energy scale), the energy level of localized
electrons $E_f$ and the chemical potentials of itinerant $\mu_d$
and localized $\mu_f$ electrons, respectively.
Here $E_f$ and $\mu_f$ are given as independent parameters,
but physical properties of the system depend only on the difference
$\mu_f-E_f$. Taking into consideration that fact, the
ground canonical phase diagram was chosen to be displayed in the variables
$\mu_d$ and $\mu_f$, as if $E_f$ would be equal to 0. However,
it is equivalent to another representation with $\mu$ and $\mu - E_f$
as independent parameters (if one puts $\mu_d = \mu_f = \mu$).

Double occupancy of the localized electrons is forbidden,
implying the on-site Coulomb repulsion $U_{ff}$ between two
$f-electrons$ is infinite.
Consequently, at a given site the $f-electron$ occupancy is assumed
to be $n_f=n_{f,\uparrow} + n_{f,\downarrow} \leq 1$ and the
$d-electron$ occupancy to be $n_d=n_{d,\uparrow} + n_{d,\downarrow} \leq 2$.
So there are 3 states per site allowed for the $f-electrons$
($n_f=0$; $n_{f,\uparrow}=1$ and $n_{f,\downarrow}=0$;
$n_{f,\uparrow}=0$ and $n_{f,\downarrow}=1$)
and 4 states per site allowed for the $d-electrons$
($n_d=0$; $n_{d,\uparrow}=1$ and $n_{d,\downarrow}=0$;
$n_{d,\uparrow}=0$ and $n_{d,\downarrow}=1$; $n_d=2$).

The first three terms of the Hamiltonian (\ref{ham}) form the
multicomponent FKM studied recently in the $2D$ case \cite{Jedrzejewski}
and in the limit of infinite dimensions \cite{ZlaticFreericksLemanskiCzycholl,FreericksZlaticRMP}.
The last term of (\ref{ham}) describes a simplified interaction that couples
spins of an itinerant and localized electron occupying the same site.
The inclusion of this interaction enables one to describe magnetic structures
(in addition to charged ones) produced by localized electrons and,
at the same time, the band magnetism of itinerant electrons.

All single-ion interactions included in (\ref{ham}) preserve states
of the localized electrons, i.e. the itinerant electrons traveling through
the lattice change neither occupation numbers nor spins of the localized
ones. In other words, $[H,f^+_{i\eta }f_{i\eta }]=0$ for all $i$ and
$\eta $, so the local occupation number is unchanged.
This is a characteristic feature of the Falicov-Kimball based models,
which makes them tractable in a controllable way.

The localized electrons play the role of an external, charge and spin
dependent potential for the itinerant electrons. This external potential
is "adjusted" by annealing, so the total energy of the system
attains its minimum. So there is a feedback between the subsystems
of localized and itinerant electrons, and this is the feedback
that is responsible for the long-range ordered  arrangements of the localized
ones, and consequently for the formation of various charge and/or spin
distributions in low temperatures.

On the other hand, if the total magnetization of the localized
electrons is non-zero, the potential experienced by the itinerant
spin-up electrons differs from that of spin-down ones. As a result,
the spins of the itinerant electrons become partially polarized,
so we also have unsaturated band magnetism.

So far it was a common practice to investigate the isotropic,
Heisenberg-type interaction between spins of localized and itinerant
electrons (e.g. within a framework of the {\em s-d} model \cite{Vonsovsky}).
Here, instead, we propose an Ising-type coupling between the spins
(see also Ref. \cite{FrohlichUeltschi}). An advantage of the
latter approach is that the model can be treated rigorously.
However, one can also provide a plausible physical background that
justifies the assumption that the spin-flip processes generated by
$S^+S^-$ and $S^-S^+$ operators, which are characteristic for the
Heisenberg-type interaction, can be neglected in a first
approximation. The reasoning comes from a simple notice that
magnetic structures observed in many materials are stable.
So, presumably, electrons moving through a crystal preserve
their spins over many lattice sites.
Besides, the Ising type coupling between itinerant electrons
was already studied in the framework of the $s-d$ model \cite{Letfulov}.

Here we show that the model (\ref{ham}) captures driving mechanismes
of formation of stripe phases and other charge and/or magnetically
ordered superstructures. The class of materials fitting this picture
is quite big and encompasses conductors that are magnetically ordered
in low temperatures, but with no sign of the Kondo effect
(as the latter implies the spin-flip process).
There is a huge number of such systems, including many compounds
of lanthanides and actinides, where various complicated magnetic
structures were detected \cite{SantiniLemanskiErdos}.
However, the model we propose is primarily aimed
to desrcibe systems that display both a charge and magnetic order.
In particular, we expect that dopped systems with a large crystal field
effect would be the best candidates.
The dopped systems, as they are usually composed of ions with different
occupancies of localized electrons (a mixed-valency regime) and because they
usually contain band electrons.
This category includes materials, where stripe phases were recently
detected as e.g. $La_{1.6}Nd_{0.4}Sr_xCuO_4$ \cite{Tranquada},
$YBa_2Cu_3O_{6+x}$ \cite{MookDaiDogan}, $Bi_2Sr_2CaCu_2O_{8+\delta }$
\cite{HowaldEisakiKanekoKapitulnik} or $La_{1.5}Sr_{0.5}NiO_4$
\cite{KajimotoIshizakaYoshizawaTokura}.
The behaviour of systems like those was already analyzed theoretically
on a basis of the extended Hubbard model
\cite{ZaanenOles,GoraRosciszewskiOles}
and the $t-J$ model \cite{HellbergManousakis}.

On the other hand, a strong crystal field makes
a flip of magnetic moment of an ion difficult. Consequently, a process of
a simultaneous flip of spins of a localized and an itinerant electron becomes
rare. This is another justification of taking into account only the Ising-type,
instead of isotropic Heisenberg-type coupling in the model.

A major interaction omitted in the Hamiltonian (\ref{ham})
is the Hubbard-type interaction between spin-up and spin-down
electrons. An inclusion of that term would make it intractable
rigorously for arbitrary values of the coupling constants,
but once again, one can imagine a simple justification
for the omission. Namely, it is assumed here that the strength
of the on-site interactions between two particles abides by
the following hierarchy:
a) the largest one - when the both particles are localized,
b) an intermediate one - if one of the particles is localized
and the other is itinerant,
c) a negligible one - when the both particles are itinerant.
This hierarchy may be summarized by a rather natural rule:
the longer time particles occupy the same site, the more
important becomes interaction between them. We point out here
that the cases a) and b) are treated exactly within the model we propose.
Obviously, there are no obstacles to include various interactions
neglected in the Hamiltonian (\ref{ham}) as perturbations.
On the other hand, the simplicity of the model (\ref{ham})
makes it attractive for studies using various techniques,
both analytical and numerical.

An influence of the Hubbard interaction $U_{dd}$ between spin-up and spin-down
itinerant electrons on an arrangement of the localized
electrons is not known in a general case, but for $U_{dd}$ small
we don't expect any dramatic changes of the phase diagram.
However, if $U_{dd}$ is large, the situation may be different.
For example, for half-filling (one itinerant electron per site)
the itinerant electrons are ordered aniferromagnetically
in the large $U_{dd}$ limit and they impose the same simple
type of order to the localized electrons.
Then, if there is one localized electron per site, the
configuration of the localized electrons will be antiferromagnetic,
i.e. the same as in the case studied in the current paper.
If, however, the density of localized electrons is equal to 1/2,
the ground-state configuration D3 (see Fig. 2c) of the model
(\ref{ham}) will have higher energy than the phase corresponding
to the configuration $b$ displayed in Fig. 2a. So, if the Hubbard
interaction is taken into account transformations of ground-state
configurations of localized electrons are expected for some pairs
of densities of electrons ($\rho_d,\rho_f$) but not for the whole
range of their values.

In the current paper we study the ground-state phase diagrams
of the model (\ref{ham}) on the square lattice.
However, the method we use can be applied to various
types of $one$- $two$- or $three-dimensional$ lattices.
The studies are based on the method of restricted phase diagrams
\cite{WatsonLemanski} constructed in the grand canonical ensemble
(in the plane ($\mu_d$, $\mu_f$) - see Fig. 1a) and then translated
into a canonical ensemble diagram (in the plane
($\rho_d$, $\rho_f$) - see Fig. 1b). Working within the framework
of a grand canonical ensemble first assures the thermodynamic
stability conditions are fulfilled
(see Ref. \cite{GajekJedrzejewskiLemanski}).

Another important reason for using the chemical potentials
$\mu_d$ and $\mu_f$ in the model (\ref{ham}) is a possibility
of changing and adjusting the electron's occupation numbers.
Then, although the model does not contain a hybridization term, suitable
changes of the chemical potentials or the position of the $E_f$ level can
produce appropriate changes of the occupation numbers. In particular,
fixing a total number of electrons enables us to study classical intermediate
valence states, where some localized electrons change their occupancy to move
into the delocalized states, conserving the total number of electrons.
Of course, a change of the chemical potentials or $E_f$ cannot evoke
a change of the electron states but only their occupation numbers.

Previous work on the spinless FKM has revealed a rich structure
of the phase diagrams \cite{WatsonLemanski,LemanskiFreericksBanach}.
A number of charge ordered superstructures has been found, including
stripe and nonstripe phases. However, so far little is known
about magnetic structures accompanying those of charge ones,
except for the limit of large $J$, where the ferromagnetic order was proven
to exist for $\rho_d<\rho_f\leq 1$ \cite{FrohlichUeltschi}.
As far as we know, the present study represents the first analysis
of both charge and magnetic ordering in the framework of a
generalized Falicov-Kimball based model for finite $U$ and $J$ values.

In the next section we explain how the calculations were carried out.
The restricted phase diagrams are presented and described in
the section III. The last section contains summary and conclusions.

\section{Method of calculation}
Here we consider all possible configurations
of the localized $f-electrons$ (including their spins), for which
the number of sites per unit cell $N_0$ is less or equal to $N_c=4$.
Taking into account all allowable sizes and shapes
of the unit cells, as well as all relevant translational vectors,
we found 47 distinct configurations.
Some of them, representing phases with the smallest periods
(up to 2 sites in an unit cell) and those with higher periods that
are analyzed in this paper, are displayed in Fig. 2. The presentation
of the configurations is chosen in such a way that make an easy observation
of their transformations, starting from the ferromagnetic phase (F) and ending
either at the antiferromagnetic one (AF) (Confs. 1--4), or at the "empty"
one (E) (Confs. D1--D5), in accordance with the diagrams given in Fig. 1.

For each periodic configuration in our trial set we performed the Fourier
transformation of the Hamiltonian (\ref{ham}) and determined the electronic
band structure for the conduction electrons. In other words, we solved
the eigenvalue problem and found the eigenvalues $E_{\nu \sigma k}$,
with branch index $\nu =1,2,...,N_0$, spin index $\sigma $
and the Bloch wavevector $k=(k_x,k_y)$
(for more details see Refs. \cite{WatsonLemanski,LemanskiFreericksBanach}).
This required us to diagonalize up to $4\times 4$ matrices and resulted
in analytical formulae for at most 4 different energy bands,
separately for spin-up and spin-down electrons.
(It is the main reason why the maximum size of the unit cell
is limited to $N_c=4$ only, as the analytical formulae for phases
with larger unit cells are not known in the general case.)
In the simplest cases, related to 6 configurations
with unit cells containing 1 or 2  sites
(Confs. E, F, AF, a, b -- see Fig. 2a and Conf. 3 -- see Fig. 2b),
the eigenvalues are given in the Appendix. For the remaining
configurations, with unit cells containing 3 or 4 sites,
the analytical formulae are quite long and for that reason
are not displayed here.

Having exact formulae for the energy spectra $E_{\nu \sigma k}$ all
quantities of interest can be calculated from the densities of states
of spin-up and spin-down electrons
\begin{equation}
Z_{\sigma }(E)=\sum_{\nu }^{}{\int_{BZ}^{}{dk\delta (E-E_{\nu \sigma k})}}.
\end{equation}
In particular, the electron densities
\begin{equation}
\rho _{d\sigma }(\mu _d)=\int_{-\infty }^{\mu _d}{Z_{\sigma }(E)dE}
\end{equation}
and the total electronic energy per site
\begin{equation}
E_{tot}(\mu _d)=\int_{-\infty }^{\mu _d}{E(Z_{\uparrow}(E)+Z_{\downarrow}(E))dE}
\end{equation}
as a function of the chemical potential (the Fermi energy) $\mu _d$.
The energy per site in the ground canonical ensemble
(the Gibbs thermodynamical potential) is defined as
\begin{equation}
E_{gc}=E_{tot}-\mu _d(\rho _{d\uparrow}+\rho _{d\downarrow})-
\mu _f(\rho _{f\uparrow}+\rho _{f\downarrow}).
\end{equation}
For the simplest configurations E, F and AF the quantities can be obtained
analytically, but the expressions are rather complicated
(they are given in terms of special functions).
This is way in our calculations we employ a Brillouin zone grid of $100\times 100$ momentum
points for each bandstructure. At each discrete momentum point of the Brillouin zone
we get at most 4 eigenvalues. Then the eigenvalues of the bandstructure are summed
to determine the ground-state energy for each number of conduction electrons.
The Gibbs thermodynamical potential is calculated for all possible values
of the chemical potentials of the conduction and localized electrons through the formula
\begin{equation}
E_{gc}(\left\{ w_i \right\})=
\sum_{\sigma }(\frac{1}{10000N_0}\sum_{\epsilon _{j\sigma} <\mu _d}{\epsilon _{j\sigma }
(\left\{ w_i \right\})-\mu _d\rho _{d\sigma }-\mu _f\rho _{f\sigma }}),
\end{equation}
where $w_i=0$ or $\uparrow$ or $\downarrow$ characterizes the localized electron
state at site $i$, and the symbol $\epsilon _{j\sigma }(\left\{ w_i \right\})$
denotes the energy eigenvalues of the bandstructure for the given configuration
of localized electrons $\left\{ w_i \right\}$.
The ground canonical phase diagram is constructed in the plane of
$(\mu_d, \mu_f)$. We directly compare
the ground state energies of all phases from the trial set, and select
the lowest one. Finally, the grand-canonical diagram is translated to a canonical
phase diagram for arbitrary densities $\rho_f=\rho _{f\uparrow}+\rho _{f\downarrow}$
and $\rho_d=\rho _{d\uparrow}+\rho _{d\downarrow}$ of the $f-$
and $d-electrons$, respectively. This procedure assures
thermodynamical stability of all phases (both periodic and their
mixtures) present in the resulting canonical phase diagrams
\cite{GajekJedrzejewskiLemanski}.

\section{Results and discussion}
The grand canonical phase diagrams plot phase boundaries, which divide
the plane of chemical potentials into domains, in each of which a single
localized electron configuration forms the ground state. The general
structure of the diagrams (Fig. 1a) is similar to that found for
the spinless FKM \cite{WatsonLemanski}. It is relatively simple for
large $U$ and more complex for small $U$. However, since the spin
degeneracy is now lifted, new domains related to various magnetic
arrangements having the same distribution of localized particles appear.
In particular, the upper region corresponding to the full configuration
in the spinless case is now split into a set of phases, ranging from
F, through ferri- and various complex antiferromagnetic
to the simplest AF phase; see Figs. 1, 2a,b.
The lower region corresponds to the empty configuration (E -- see Fig. 2a),
and the largest of the remaining regions is the diluted antiferromagnetic
phase (D3 -- see Fig. 2c).

In the case of $U = 8$, $J = 0.5$ the central portion of the diagram
is divided into diagonal stripes in which various periodic configurations
of localized electrons are ground states. The sequence of the localized
electron densities, reading from the left to right at fixed $\mu_f$ is 3/4,
2/3, 1/2, 1/3 and 1/4 (Confs. D1--D5, see Fig. 2c), the ${\rho_f = 1/2}$
phase (D3) being the largest region.

In the corresponding canonical diagram (in the plane $({\rho_d, \rho_f})$ )
given in Fig. 1b all these phases are situated along the line
${\rho_f + \rho_d/2 = 1}$. Movement along that line (starting from F)
corresponds to a dilution of the ferromagnetic state by a gradual replacement
of the localized electrons by itinerant ones in such a way, that each localized
electron is replaced by two itinerant ones.
This is different from the spinless case, where periodic phases are located
along the line ${\rho_f + \rho_d = 1}$. The difference is related to the fact,
that now both spin up and spin down itinerant electrons can independently occupy
each site, so the total density $\rho_d$ can take any value between 0 and 2.

The analysis of the canonical phase diagram displayed in Fig.1b shows a strong
tendency to antiferromagnetic order in the region of densities ${\rho_d}$
around 1 (half-filling) and to the ferromagnetic order outside that region.
Indeed, both the phases D2 and D3 (see Fig. 2c) situated along the line
${\rho_f + \rho_d/2 = 1}$ and those represented by the configurations
no. 3, 4 and AF (see Figs. 2a,b) situated along the line ${\rho_f = 1}$
are ordered antiferromagnetically.

For intermediate values of the density ${\rho_d}$ and ${\rho_f = 1}$
both antiferro- and ferrimagnetically ordered phases are found
(Conf. 1 and 2 -- see Fig. 2b).
These phases represent consecutive stages of a reconstruction process
when F transforms into the simplest AF with an increase of ${\rho_d}$.

On the other hand, only ferro- (F, D1, D4, D5 -- see Figs. 2a,c) and
antiferro- (D2, D3 -- see Fig. 2c), but not ferrimagnetically ordered phases
were found along the line ${\rho_f + \rho_d/2 = 1}$. Moving along that line
(in Fig. 1b) from the left-upper corner ($\rho_d=0$,$\rho_f=1$) to right-lower
one ($\rho_d=2$,$\rho_f=0$) one can notice that the process of dissolution
of the full ferromagnetic phase overlaps with the tendency
to antiferromagnetism inside a region around half-filling.

The picture becomes more complicated as $U$ is reduced. An example is the case
of $U = 2$ and $J = 0.5$ (not displayed here), where complex
structures of the diagrams are found both in the plane of chemical
potentials $(\mu_d, \mu_f)$ and densities $({\rho_d, \rho_f})$.
First of all, more phases appear in the diagrams (20, versus 12 in the case
of $U = 8$) and the domains occupied by the phases in the central region
are no longer diagonal stripes parallel to each other, but
have a less regular structure.

The phases from the central region are distributed in different sectors
of the plane $({\rho_d, \rho_f})$ in the corresponding canonical phase
diagram. Only three periodic phases are found along the line
${\rho_f + \rho_d/2 = 1}$: D3, D5 and D1a -- see Figs. 2c,d.
Two of them (D3 and D5) already appeared in the $U=8$ case, but the
ferrimagnetically  polarized D1a phase replaces the ferromagnetic D1
(compare Figs. 2c,d). This fact illustrates a possibility of phase
transitions from one magnetic structure to another with a change of $U$.

Another characteristic feature of the periodic phases found in the $U = 2$
diagram is their stability with respect to finite intervals
of the density ${\rho_d}$. This means that the Fermi level of the
corresponding phase lies inside its energy band, so those phases are conducting.
This is just the opposite to the situation found in the $U = 8$ diagram
(Fig.1b), where the phases D1--D5 are ground states
for fixed values of itinerant electron densities ${\rho_d}$,
corresponding to insulating states (the Fermi levels
are situated inside their energy gaps).

Presumably, if we increase the maximum size of unit cells $N_c$, then phases
with larger periods will enter the phase diagram in such a way, that intrevals
of the density ${\rho_d}$ where low-period phases are stable will decrease.
It is not clear from our studies if the intervals shrink to single values
or to finite intervals of the density in the full phase diagram.
Perhaps the first scenario occurs for some phases and the other for the rest.

A common feature of the canonical diagrams for $U = 8$ and $U = 2$
is that in both cases the fully occupied phases with ${\rho_f=1}$
(Confs. F, 1--4, AF) are the same and have identical positions.
It appears that a distribution of these phases is symmetric
with respect to half-filling ${\rho_f = 1}$
and it does not depend on $U$, but merely on $J$.

Phases with ${\rho_f = 1}$ that appear in the diagrams shown in Fig. 1
seem to be stable within finite intervals of ${\rho_d}$. This means
that  their ground states can be conducting. In particular,
antiferromagnetic  conductors appear to be possible ground states
for certain electron  concentrations off of half-filling.
However, this is probably an artifact due to the restriction imposed
by the maximum period of configuration in the trial set.
If one takes into account configurations with larger periods then,
presumably, some of them will enter the phase diagrams in such a way,
that more and more phases will be stable merely within energy
intervals lying inside one of their energy gaps. Therefore, one expects that
the full phase diagram will contain many periodic phases in the insulating
states, and only a part of them, especially those having gapless energy spectra,
as e.g. those represented by the Conf. E and F, will be conducting.

A similar situation was observed for the spinless FKM, where large
period phases occupy regions of the phase diagram
located outside of the energy gaps of phases having low periods
\cite{LachLyzwaJedrzejewski,WatsonLemanski}.

\section{Summary and conclusions}
A simple model capable of describing both charge and magnetic structures
of localized electrons, as well as itinerant band magnetism, was introduced
and investigated on a square lattice. Restricted ground
state phase diagrams were constructed for intermediate values of the coupling
parameters. Various types of charge and magnetically ordered phases were
detected for a range of band fillings, illustrating e.g. consecutive
stages of transformation of F to AF with an increase of the band filling.

It is remarkable, that the results presented here are consistent with those
found for the Hubbard model in the large $U$ limit \cite{Anderson} and also
with those obtained in Ref. \cite{FrohlichUeltschi}, where - in turn - the
limit of large $J$ was investigated, as in all these cases one gets
the simplest AF at half-filling and F far away from that limit.
However, according to Ref. \cite{FrohlichUeltschi} and Ref. \cite{Anderson}
the ferromagnetic ground state extends to all band fillings but half-filling,
which is consistent with the Nagaoka theorem \cite{Nagaoka}. This, of course,
is a consequence of investigation of the large $U$ or $J$ limit.

Here, instead, we provide results that give an opportunity to determine
the upper limit for the density of itinerant electrons below which  F can be
stable for finite $U$ and $J$. Indeed, a direct comparison of energies of phases
from the restricted set (they were calculated rigorously by using the exact
analytical formulae for their energy spectra) shows that F can be stable  for
$\rho_d\leq 0.131$ (or $\rho_d\geq 2 - 0.131$) if $J=0.5$ and for
$\rho_d\leq 0.233$ (or $\rho_d\geq 2 - 0.233$) if $J=1$. These results complete
those given in Ref.\cite{FrohlichUeltschi}, where no critical value for $\rho_d$
below which F is stable was reported. From the theorems  provided in
Ref. \cite{FrohlichUeltschi} it may be merely concluded, that F is stable
for any $0<\rho_d < \rho_f$ if $\rho_f=1$ and $J$  is large enough.

Since no restrictions were here imposed on values of $U$ or $J$, it was not
a great surprise that other ordered phases than just the simplest AF and F
have been detected for intermediate electron concentrations. In particular,
various magnetically ordered phases were found for $\rho_f=1$ (when charge
is distributed uniformly) and intermediate $U$ and $J$, illustrating
a transformation of F to AF with an increase of band filling.

But what are the physical reasons and what are the driving forces that lead to the complex
charge and/or magnetic arrangements? If one attempts to describe such behavior
in terms of two-body forces, then interactions between more distantt than
the nearest neighboring lattice sites needs to be exploited. The double exchange
\cite{ZenerAnderson} and RKKY interactions, derived from the correlated electron
models by using perturbative methods, are prominent examples of such approaches.

On a basis of the present studies we conclude that the phenomenon of charge
and/or magnetic ordering can be explained by a mutual adjustment of the distributions
of the charges and spins of the itinerant and localized electrons. This idea is
not new, since it constitutes a background for the famous density functional
theory. The difference is that we already start from an effective Hamiltonian
(represented here by (\ref{ham})) and consider many lattice sites instead of
focusing on the details of a charge distribution around a single site, as is
commonly practiced in the {\em ab initio} calculations. This allows us to take
into account the kinetic energy of the itinerant electrons and to extract
the most essential information about the system under investigation.

Indeed, one needs to investigate large enough areas accessible for itinerant
electrons, as their total energy depends on the distribution of the localized
ones over many lattice sites. If the latter forms a charge and/or magnetically
ordered structure, then the mean values of the charge density distribution
and/or the spin polarization of the moving electrons also adopts a suitable
structure. It is clear that in the simplest case of a ferromagnetic metal,
the densities of itinerant spin-up and -down electrons spread out uniformly
over the whole lattice, but their values on each site are different,
i.e. the moving electrons are polarized. But in the general case
the distributions of itinerant spin-up and -down electrons are
inhomogeneous and they may differ one from another.

Such an inhomogeneous distribution means that the itinerant electrons have some
more preferable routes of travelling though the crystal (the routes may be
different for the spin-up and -down electrons). So one can notice here
something like a traffic self-regulation, where the system of localized
electrons orders in such a way that the itinerant electrons would have
as much freedom to move as possible. This of course is governed by the
quantum mechanical laws through a minimization procedure of the total
energy, which depends not only on the coupling constants, but also
on the densities of the electrons.

It is interesting that many of the stable structures
are axial (the lines of equivalent sites are parallel
to the lattice axis) or diagonal (the corresponding lines
are oriented along the (1,1) direction) stripes. Axial stripes are
predominant for rather small band fillings, i.e. for small densities
of itinerant electrons, and the diagonal stripes are found close to the
half-filling. The same picture was already observed for the
spinless FKM \cite{LemanskiFreericksBanach}.

This stripe-type ordering means that the itinerant electrons prefer to move
along simple, one-dimensional channels. Has this observation anything to do
with, or can it shed light on a mechanism of the $high-T_c$
superconductivity? It is too early to
answer this question, but the fact is that the stripe ordered phases
have been detected in many $high-T_c$ materials. So further studies
along the lines indicated here will be interesting.

Another point is that the detection of charge and magnetic phases other
than F or AF provides an opportunity to describe various structures
observed in many systems. Indeed, for many years various complex magnetic
structures were observed in materials containing transition metals,
rare-earths  or actinides. Recently, thanks to new precise experimental
techniques, stable charge superstructures have been also found in a number
of compounds (see e.g. \cite{MookDaiDogan} and the references given there).
The model proposed here is capable to describe such structures in a simplified
way but, of course, this work is merely the first step towards a complete
analysis of the apparently very complicated processes occurring in real materials.

It is worthwhile to notice, that even though only the restricted phase diagrams
were analyzed here, the conclusions are expected to also hold for the complete
phase diagrams. This conjecture comes from the fact that the reported
results are consistent with those obtained exactly in limiting cases and
from a comparison with the data found previously for the spinless FKM,
where an increase in the size of the allowed unit cells does not produce
significant qualitative change in the phase diagram.

In closing we admit that it is clear that the simple effective
model presented in this paper cannot describe many interesting phenomena
observed in solids, like the Kondo effect or the superconductivity.
However, since it is already a non-trivial model that can describe some
of the phenomena, it can serve as a reference system for studies of more
elaborate models of correlated electron systems.

\newpage

\acknowledgements
I would like to thank Prof. G. Czycholl, Prof. P. Fulde, Dr. E. Runge,
Prof. J. Spa{\l }ek, Dr. J. Ulner, Dr. P. Wr\'obel and Dr. A. N. Yaresko
for useful discussions on the topics related to this paper
and prof. J. K. Freericks for helpful remarks and critical reading
of the manuscript.
This work was partially supported by
The Max Planck Institute for Complex Systems in Dresden, Germany.

\appendix

\section{}
Below are given exact formulae for the energy spectra $E_{jk\uparrow }$ of the spin-up
itinerant electrons in the simplest cases of period 1 and 2 configurations
for the localized electrons (Conf. \textit{E, F, AF, 3, a, b} -- see Figs. 2a,b).
The corresponding spectra of the spin-down electrons can be obtained by taking
the opposite signs of parameter $J$.
\begin{eqnarray}
E_{1k\uparrow }^E(k_x,k_y)&=&2(cosk_x+cosk_y)\\
E_{1k\uparrow }^F(U,J,k_x,k_y)&=&U-J+2(cosk_x+cosk_y)\\
E_{^1_2k\uparrow }^{AF}(U,J,k_x,k_y)&=&U\mp \sqrt{J^2+4(cosk_x+cosk_y)^2}\\
E_{^1_2k\uparrow }^3(U,J,k_x,k_y)&=&U+2cosk_y\mp \sqrt{J^2+4(cosk_x)^2}\\
E_{^1_2k\uparrow }^a(U,J,k_x,k_y)&=&\frac{1}{2}\left[U-J+4cosk_y\mp \sqrt{(U-J)^2+16(cosk_x)^2}\right]\\
E_{^1_2k\uparrow }^b(U,J,k_x,k_y)&=&\frac{1}{2}\left[U-J\mp \sqrt{(U-J)^2+16(cosk_x+cosk_y)^2}\right]
\end{eqnarray}

\newpage

\begin{figure}
\caption{The restricted ground-state phase diagrams for $U = 8$ and $J = 0.5$:
(a) the ground canonical phase diagram, i.e. in the plane of chemical potentials
($\mu_d$, $\mu_f$),
(b) the canonical phase diagram, i.e. in the plane of densities ($\rho_d$, $\rho_f$).
F, AF and E refer to the ferromagnetic, antiferromagnetic
and empty (i.e. without localized f-electrons) phases, respectively.
Periodic phases with $\rho_f=1$ are denoted by the numbers 1--4 (see Fig. 2b),
and those with ${\rho_f + \rho_d/2 = 1}$  by the symbols D1--D5 (see Fig. 2c).
(The domains D2 and D4 are very thin, therefore their boundaries are so close
to each other that they have the appearence of a single thick line.)
In the case (a), the phases occupy finite size domains
and mixtures refer to the lines that separate the domains.
In the case (b), the phases are ground states only on the bold straight-line segments
or at single points. Outside these straight line segments or points
there are mixtures of periodic phases that have lower energy than any
periodic phase taken from the restricted set.
The small vertical straight line segments crossing the $\rho_f=1$ line mark
limits of F.
The diagonal line ${\rho_f + \rho_d/2 = 1}$ is only a visual guide.}
\label{fig1}
\end{figure}

\begin{figure}
\caption{Pictures of configurations of the localized electrons representing
phases discussed in the paper. The ${\uparrow}$ (${\downarrow}$) refer
to sites occupied by a localized electron with spin up (down) respectively,
and the dots denote sites not occupied by the localized electrons. The shaded
region in the lower left corner shows the unit cell, and line  segments show
the translation vectors that are used to tile the two-dimensional plane.
Panel (a) shows the simplest configurations of the localized electrons containing 1 or 2
sites per unit cell. Three of them (E -- empty, F -- ferromagnetic and
AF -- antiferromagnetic) appear in the phase diagrams and the last two
($a$ and $b$) are not present in the diagrams.
Panel (b), the configurations 1--4 with $\rho_f=1$, which are found
on the restricted phase diagrams for $U=8$ and $J=0.5$
between F and AF.
Panel (c), the configurations D1--D5 with $\rho_f < 1$, which are found in the
restricted phase diagrams for $U=8$ and $J=0.5$ between
the F and E phases; on the canonical phase diagram (Fig. 1b)
they appear along the diagonal $\rho_f + \rho_d/2 = 1$.
Panel (d), the configurations with $\rho_f = 1$ (Confs. 1a, 5), that are
found between the F and AF phases in the restricted phase diagrams for $U=8$
and $J=1.0$, but are not found on the diagrams for $U=8$ and $J=0.5$:
Conf. 1a replaces Conf. 1, whereas Conf. 5 appears between
Conf. 4 and AF. The configuration D1a with $\rho_f = 0.75$ is found
on the restricted phase diagrams for $U=2$ and $J=0.5$ instead of D1
(see Fig. 2b).}
\label{fig2}
\end{figure}


\newpage
\begin{references}

\bibitem{FalicovKimball}
	L. M. Falicov and J. C. Kimball,
	Phys. Rev. Lett. {\bf 22}, 997 (1969).
\bibitem{LachLyzwaJedrzejewski}
	J. Lach, R. \L y\.{z}wa and J. J\c{e}drzejewski,
	Acta Phys. Pol. A {\bf 84}, 327 (1993);
	Phys. Rev. B {\bf 48}, 10783 (1993).
\bibitem{WatsonLemanski}
	G. I. Watson and R. Lema\'{n}ski
	J. Phys. Condens: Matter {\bf 7}, 9521 (1995).
\bibitem{LemanskiFreericksBanach}
	R. Lema\'{n}ski, J. K. Freericks and G. Banach,
	Phys. Rev. Lett. {\bf 89}, 196403 (2002);
	J. Stat. Phys. {\bf 116}, 699 (2004).
\bibitem{Brandt}
	U. Brandt, A. Fledderjohann and G. H\"ulsenbeck,
	Z. Phys. B {\bf 81}, 409 (1990);
	U. Brandt and A. Fledderjohann,
	Z. Phys. B {\bf 87}, 111 (1992).
\bibitem{Farkasovsky}
	P. Farka\v{s}ovsk\v{y},
	Phys. Rev. B {\bf 54}, 11261 (1996).
\bibitem{FreericksZlatic}
	J. K. Freericks and V. Zlati\'c,
	Phys. Rev. B {\bf 58}, 322 (1998).
\bibitem{ZlaticFreericksLemanskiCzycholl}
	V. Zlati\'c , J. K. Freericks, R. Lema\'nski and G. Czycholl,
	Phil. Mag. B {\bf 81},1443 (2001);
\bibitem{FreericksZlaticRMP}
	J. K. Freericks and V. Zlati\'c, Rev. Mod. Phys. {\bf 75},
	1333 (2003).
\bibitem{Jedrzejewski}
	J. J\c{e}drzejewski and V. Derzhko,
	Physica {\bf A 317}, 227 (2003).
\bibitem{LemanskiWojtkiewicz}
	R. Lema\'{n}ski and J. Wojtkiewicz,
	phys. stat. sol. (b) {\bf 236}, 408 (2003).
\bibitem{MookDaiDogan}
	H. A. Mook, P. Dai and F. Do\u{g}an,
	Phys. Rev. Lett. {\bf 88}, 097004 (2002).
\bibitem{Ando}
	Y. Ando, K. Segawa, S. Komiya and A. N. Lavrov,
	Phys. Rev. Lett. 88, 137005 (2002).
\bibitem{HowaldEisakiKanekoKapitulnik}
	C. Howald, H. Eisaki, N. Kaneko, M. Greven and A. Kapitulnik,
	Phys. Rev. B {\bf 67}, 014533 (2003).
\bibitem{FrohlichUeltschi}
	J. Fr\"ohlich and D. Ueltschi,
	cond-mat/0404483.
\bibitem{Vonsovsky}
	S. V. Vonsovsky, Magnetism, New York, Wiley, 1974.
\bibitem{Letfulov}
	B. M. Letfulov and J. K. Freericks,
        Phys. Rev. B {\bf 64}, 174409 (2001).
\bibitem{SantiniLemanskiErdos}
	P. Santini, R. Lema\'{n}ski and P. Erd\"os,
	Adv. in Phys. {\bf 48}, 537 (1999).
\bibitem{Tranquada}
	J. M. Tranquada, B. J. Sternlieb, J. D. Axe, Y. Nakamura
	and S. Uchida,
	Nature {\bf 375}, 561 (1995).
\bibitem{KajimotoIshizakaYoshizawaTokura}
	R. Kajimoto, K. Ishizaka, H. Yoshizawa and Y. Tokura,
        Phys. Rev. B {\bf 67}, 014511 (2003).
\bibitem{ZaanenOles}
	J. Zaanen and A. M. Ole\'s,
	Ann. Physik {\bf 5}, 224 (1996).
\bibitem{GoraRosciszewskiOles}
	D. G\'ora, K. Ro\'sciszewski and A. M. Ole\'s,
	Phys. Rev B {\bf 60}, 7429 (1999).
\bibitem{HellbergManousakis}
	C. S. Hellberg and E. Manousakis,
	Phys. Rev. Lett. {\bf 83}, 132 (1999).
\bibitem{GajekJedrzejewskiLemanski}
	Z. Gajek, J. J\c{e}drzejewski and R. Lema\'{n}ski,
	Physica {\bf 223 A}, 175 (1996).
\bibitem{Anderson}
	P. W. Anderson,
	Phys. Rev.  {\bf 14}, 115 (1959).
\bibitem{Nagaoka}
	Y. Nagaoka,
	Phys. Rev. {\bf 147}, 392 (1966).
\bibitem{ZenerAnderson}
	C. Zener, Phys. Rev. {\bf 82}, 403 (1951); P. W. Anderson
	and H. Hasegawa, {\em ibid} {\bf 100}, 675 (1955).

\end{references}
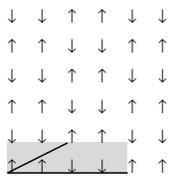
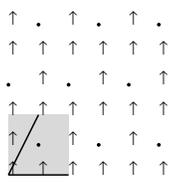
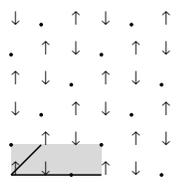
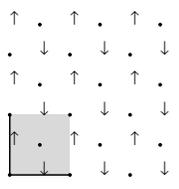
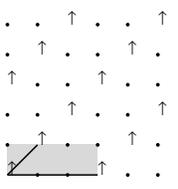
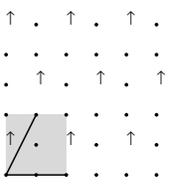
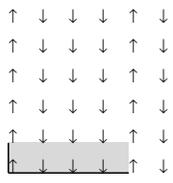
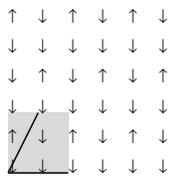
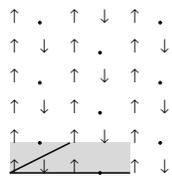
\end{document}